\documentclass[floatfix,
reprint,
superscriptaddress,
 amsmath,amssymb,
 aps,
pra,
]{revtex4-2}
\usepackage{gensymb}
\usepackage{graphicx}
\usepackage{dcolumn}
\usepackage{bm}
\usepackage{physunits}
\usepackage{xcolor}
\usepackage{amsmath}

\begin{document}

\preprint{APS/123-QED}

\title{Field-free precessional magnetization dynamics in a multidomain magnet driven by light-induced strain}

\author{Héloïse Damas}
\email{heloise.damas@ru.nl}
\affiliation{HFML-FELIX, Toernooiveld 7, 6525 ED Nijmegen, The Netherlands 
}
\affiliation{\mbox{Radboud University, Institute for Molecules and Materials, 135 Heyendaalseweg, 6525 AJ Nijmegen, The Netherlands 
}}

\author{Carl S. Davies}
\email{carl.davies@ru.nl}
\affiliation{HFML-FELIX, Toernooiveld 7, 6525 ED Nijmegen, The Netherlands 
}
\affiliation{\mbox{Radboud University, Institute for Molecules and Materials, 135 Heyendaalseweg, 6525 AJ Nijmegen, The Netherlands 
}}

\author{Tomasz Zalewski}
\affiliation{HFML-FELIX, Toernooiveld 7, 6525 ED Nijmegen, The Netherlands 
}
\affiliation{\mbox{Radboud University, Institute for Molecules and Materials, 135 Heyendaalseweg, 6525 AJ Nijmegen, The Netherlands 
}}

\author{Petr M. Vetoshko}
\affiliation{Russian Quantum Center, Skolkovo Innovation Center, Bolshoi Blv.,
30, bl.1, Moscow, 121205, Russia}

\author{Vladimir I. Belotelov}
\affiliation{Russian Quantum Center, Skolkovo Innovation Center, Bolshoi Blv.,
30, bl.1, Moscow, 121205, Russia}
\affiliation{Faculty of Physics, Lomonosov Moscow State University, Leninskie Gory, Moscow 119992, Russia}

\author{Andrzej Stupakiewicz}
\affiliation{Faculty of Physics, University of Bialytsok, 1L Ciolkowskiego, 15-245 Bialystok, Poland}

\author{Andrei Kirilyuk}
\email{andrei.kirilyuk@ru.nl}
\affiliation{HFML-FELIX, Toernooiveld 7, 6525 ED Nijmegen, The Netherlands 
}
\affiliation{\mbox{Radboud University, Institute for Molecules and Materials, 135 Heyendaalseweg, 6525 AJ Nijmegen, The Netherlands 
}}

\begin{abstract}
Ultrafast lattice excitation offers a route to controlling magnetization through magnetoelastic coupling, yet its ability to drive large-amplitude dynamics within pre-existing multidomain states remains less explored. Here, we show that a single 5-ps mid-infrared pulse induces a large-amplitude, reversible magnetic response in cobalt-doped yttrium iron garnet at zero applied magnetic field. Time-resolved magneto-optical microscopy reveals a transient reversal of the out-of-plane magnetic contrast within approximately 2 ns, followed by recovery of the initial state, while the labyrinthine domain morphology remains largely preserved. Concurrent polarization microscopy reveals a localized, long-lived strain-sensitive response and a propagating acoustic wave. The correlated structural and magnetic responses, together with micromagnetic simulations, support a mechanism in which transient strain modifies the magnetic anisotropy and drives precessional reorientation within existing domains. These findings establish single-pulse magnetoelastic excitation as a route to reversible, large-amplitude magnetization dynamics without substantial domain reorganization or an externally applied magnetic field.
\end{abstract}

\maketitle

When the magnetic equilibrium is perturbed, the resulting torque drives magnetization precession around the effective magnetic field, while damping governs its subsequent relaxation. Such precessional dynamics can lead to large-amplitude magnetization reorientation and switching, and can be initiated by short magnetic-field pulses \cite{back1999minimum,gerrits2002ultrafast}, optically induced effective fields \cite{kimel2005ultrafast}, or transient modifications of magnetic anisotropy \cite{van2002all}. The latter can arise from lattice deformation through magnetoelastic coupling, which modifies the magnetic free-energy landscape and can drive coherent precession and even magnetization switching \cite{scherbakov2010coherent,linnik2011theory,kim2012ultrafast,kovalenko2013new,thevenard2016precessional}.

Ultrafast optical excitation provides a means of generating lattice perturbations capable of driving magnetization dynamics, including coherent phonons \cite{zeiger1992theory,merlin1997generating}, propagating acoustic waves \cite{thomsen1986surface,ruello2015physical,lejman2014giant}, and quasi-static strain associated with thermal expansion \cite{shin2022quasi,walz2025large}. In dielectric crystals, mid-infrared excitation offers distinct pathways for coupling light to the lattice. Resonant excitation of infrared-active optical phonons can drive nonlinear lattice dynamics \cite{nicoletti2016nonlinear,subedi2021light,forst2011nonlinear,mankowsky2016non,disa2021engineering} and the generation of circular phonon motion \cite{davies2024phononic,fennema2026generalized}. Alternatively, excitation in the epsilon-near-zero regime can promote a more uniform deposition of energy across the sample thickness \cite{damas2026photoinduced}, generating transient thermoelastic strain \cite{kwaaitaal2024epsilon,kwaaitaal2024disentangling}. These lattice-mediated pathways have enabled the manipulation of magnetic interactions and anisotropy on ultrafast timescales \cite{afanasiev2021ultrafast,afanasiev2021controlling,stupakiewicz2021ultrafast}. However, establishing how the resulting lattice deformation drives magnetization dynamics requires resolving the structural and magnetic responses to an individual optical pulse.

Cobalt-doped yttrium iron garnet (Co:YIG) provides a suitable platform for investigating this interplay owing to its strong magneto-optical response, enhanced magnetoelastic coupling, and pronounced magnetocrystalline anisotropy \cite{hansen1977anisotropy}. Previous mid-infrared studies have demonstrated coherent magnetization dynamics \cite{frej2023laser,frej2023phonon}, phonon-driven magnetization switching \cite{stupakiewicz2021ultrafast} and labyrinthine magnetic domains reorientation in the epsilon-near-zero regime \cite{damas2026photoinduced}. These domain transformations, however, were induced by macropulses delivered by a free electron laser (FEL), which comprise approximately 200 individual 5-ps pulses separated by 40 ns, making it difficult to distinguish the elementary magnetic response from cumulative excitation effects. Whether an individual pulse can drive substantial magnetization reorientation within a pre-existing labyrinthine domain state, without inducing large-scale domain reorganization, therefore remains an open question.

Here, we investigate the magnetic and structural responses of Co:YIG to individual 5-ps mid-infrared pulses at zero applied magnetic field. Time-resolved Faraday microscopy reveals a transient reversal of the out-of-plane magnetic contrast on a nanosecond timescale, followed by recovery of the initial state, while the labyrinthine domain morphology remains largely preserved. Concurrent polarization microscopy reveals a localized, long-lived strain field and an outward-propagating acoustic wave. The correlated excitation dependences of the structural and magnetic responses, together with micromagnetic simulations, support a mechanism in which transient lattice deformation modifies the magnetic anisotropy and drives precessional magnetization reorientation within existing domains. Our results identify a reversible single-pulse response that is distinct from the persistent domain transformations observed under repeated excitation, demonstrating that large-amplitude magnetic reorientation can occur within a multidomain structure without substantial domain reorganization.\\

\begin{figure}[t]
    \centering
    \includegraphics[width=1\linewidth]{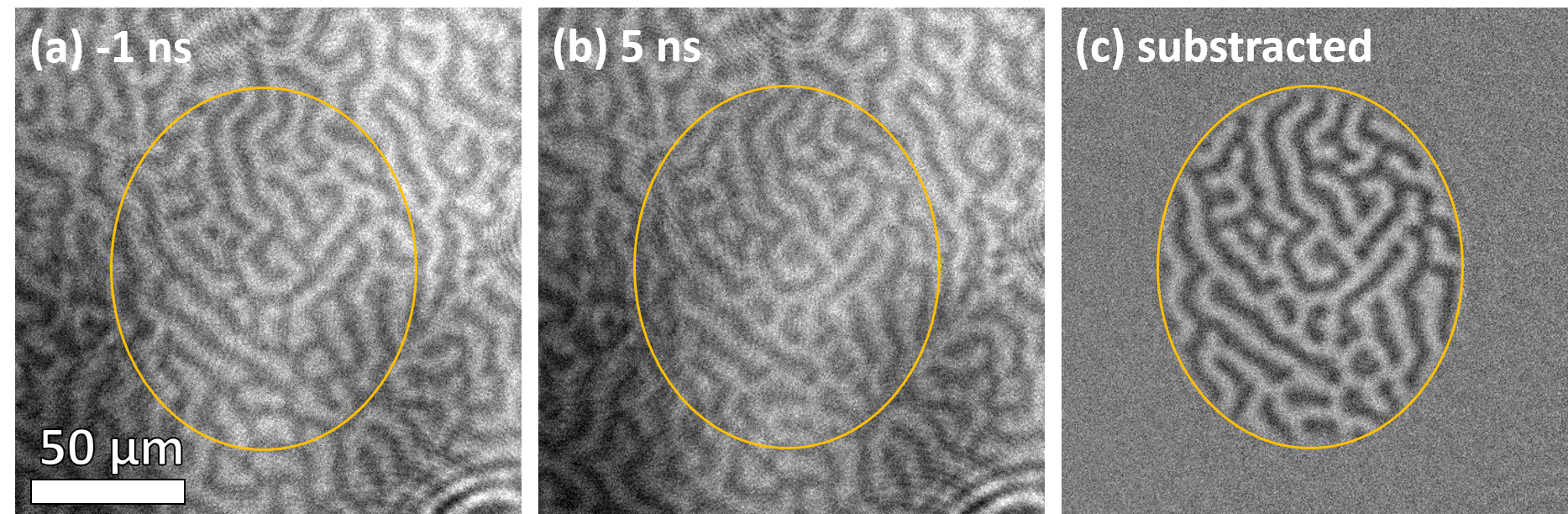}
    \caption{(a) Labyrinthine magnetic domain structure of Co:YIG before excitation. (b) Transient magnetic response visualized 5 ns after optical excitation by a single pulse ($\lambda~=~17.5~\mu$m, $E=33.8 \pm 3.5 \, \mu$J). (c) Difference image highlighting the region affected by the
    mid-infrared pulse.}
    \label{fig:mag-dyn}
\end{figure}

The experiments were performed on a 13-$\mu$m-thick film of Y$_{1.94}$Ca$_{1.22}$Fe$_{3.52}$Ge$_{1.31}$Co$_{0.14}$O$_{12}$, grown by liquid-phase epitaxy on a (001)-oriented gadolinium gallium garnet substrate. The interplay between cubic and induced uniaxial anisotropies favors equilibrium magnetization orientations close to the body diagonals of the cubic unit cell \cite{marysko1994anisotropy,maryvsko1995cubic,tekielak1997magnetic}. At zero applied magnetic field, the competition between magnetic anisotropy and magnetostatic interactions gives rise to the labyrinthine domain structure \cite{hubert2008magnetic} shown in Fig.~\ref{fig:mag-dyn}(a).

Magnetic and structural dynamics were investigated using a single-shot two-color pump-probe microscope at the FELIX facility in Nijmegen, the Netherlands \cite{oepts1995free}. Individual 5-ps mid-infrared pump pulses were isolated from the free-electron laser by cavity dumping \cite{janssen2022cavity}. The linearly polarized pulses, tunable in wavelength from 8 to 22~$\mu$m, were focused onto the sample to a spot diameter of approximately 100~$\mu$m. The out-of-plane magnetization component was monitored by time-resolved Faraday microscopy using synchronized 400-fs probe pulses at 1040~nm from a Yb-doped fiber laser. Simultaneously, time-resolved polarization microscopy was used to probe the structural response. Further details of the experimental setup and the separation of magnetic and structural contrast are provided in Supplemental Materials 1 and 2, respectively, and in Ref.~\cite{zalewski2025direct}. \\

Upon excitation with a single 5-ps mid-infrared pulse at $\lambda=17.5~\mu$m, the magnetic contrast reverses within the illuminated region (see Fig.~\ref{fig:mag-dyn}(a,b)). The magnetic contrast is reversed, while the underlying labyrinthine domain pattern retains exactly the shape and spatial distribution of the initial state, with no large-scale reorganization of the domain structure. This transient response is independent of the relative orientation between the pump polarization and the crystallographic axes of the Co:YIG crystal. At longer time delays, the initial magnetic contrast is gradually restored, highlighting the reversibility of the response. The difference image in Fig.~\ref{fig:mag-dyn}(c) highlights its spatial extent and shows that the induced dynamics are confined to the optically excited region within a sharp boundary, and predominantly follow the pre-existing domain pattern.

\begin{figure}
    \centering   
    \includegraphics[width=1\linewidth]{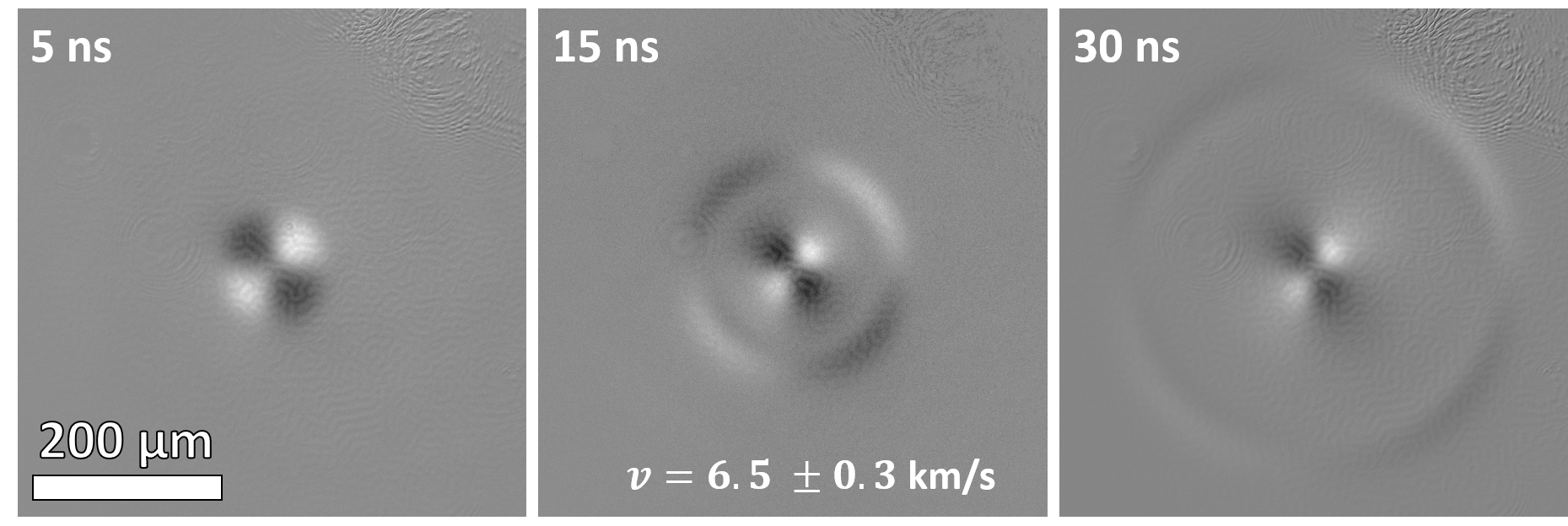}
    \caption{Time-resolved polarization microscopy images showing the
    localized strain response and outward-propagating elastic wave ($\lambda=11 ~\mu$m, $E=13.7 \pm 7.3 \, \mu$J).}
    \label{fig:strain}
\end{figure}

In parallel, time-resolved polarization microscopy captures a pronounced structural response following mid-infrared excitation (see Fig.~\ref{fig:strain}). A characteristic four-lobe pattern develops within the excited region, exhibiting alternating polarization contrast and remaining spatially localized around the pump spot. The signal intensity is independent of the relative orientation between the probe polarization and the crystallographic axes, while rotating the analyzer by 90$^\circ$ reverses the contrast. This polarization dependence confirms the birefringent origin of the signal, consistent with strain-induced birefringence following mid-infrared absorption \cite{stupakiewicz2021ultrafast,kwaaitaal2024epsilon,kwaaitaal2024disentangling}. In addition to this localized response, a second component propagates radially outward from the excitation region, forming an expanding wavefront whose displacement increases linearly with time. A propagation velocity of $v=6.5\pm0.3$ km/s is obtained from the temporal evolution of the wavefront, consistent with longitudinal acoustic velocities reported for garnet materials \cite{clark1961elastic,zhou2011pressure}. As shown in Supplemental Material 4, the strain wave velocity is independent of the pump wavelength.
The simultaneous observation of localized strain-induced birefringence and a propagating elastic wave provides direct evidence that a single mid-infrared pulse produces a substantial transient deformation of the lattice.

\begin{figure}[h!]
    \centering
    \includegraphics[width=1\linewidth]{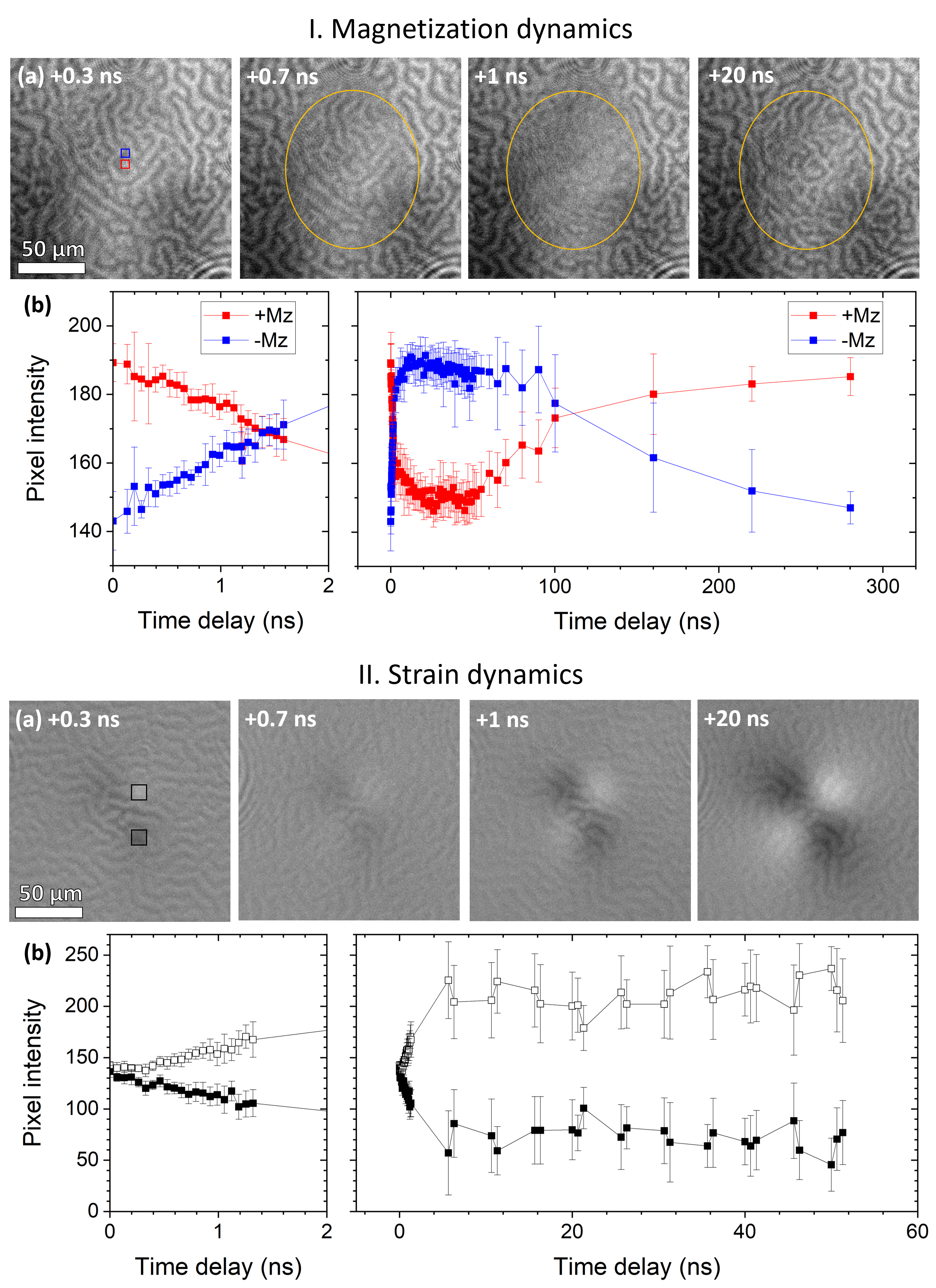}
    \caption{\textbf{I. Magnetization dynamics.} (a) Time-resolved images following excitation ($\lambda=17\,\mu\mathrm{m}$, $E=32.5\pm3.1\,\mu\mathrm{J}$). (b) Pixel intensity of domains with opposite magnetic contrast. The contrast decreases within 1 ns, reverses after $\sim1.5$ ns, and recovers within 200-300 ns. \textbf{II. Strain dynamics} (a) Time-resolved strain images ($\lambda=11,\mu\mathrm{m}$, $E=13.7\pm7.3\,\mu\mathrm{J}$). (b) Pixel intensity of lobes with opposite contrast. The strain signal increases immediately after excitation and remains nearly constant for $\sim50$ ns, consistent with a quasi-static component.
    }
    \label{fig:time dependence}
\end{figure}

The temporal evolution of the magnetic and structural responses is compared in Fig.~\ref{fig:time dependence}. In the time-resolved magnetic images (Fig.~\ref{fig:time dependence}.I(a)), a progressive reduction of the domain contrast is observed within the first nanosecond following excitation, followed by reversal of the contrast at longer delays. This evolution is quantified in Fig.~\ref{fig:time dependence}.I(b) by tracking the pixel intensity within domains of opposite initial magnetic contrast. The two intensities progressively approach each other following excitation and cross after approximately 1.5~ns, marking the reversal of the magnetic contrast. At longer delays, the initial contrast is gradually recovered, with relaxation occurring on a timescale of approximately 300~ns. The timescale of the magnetization response depends on the excitation energy, as shown by the pump-energy dependence of the magnetic contrast at a fixed pump-probe delay (see Supplemental Material 5).

The evolution of the strain response is shown in Fig.~\ref{fig:time dependence}II. Within the temporal resolution of the experiment, the four-lobe birefringence pattern is observed immediately following excitation and becomes progressively more pronounced during the first few nanoseconds (Fig.~\ref{fig:time dependence}.II(a)). This evolution is quantified by tracking the pixel intensity within lobes of opposite contrast (Fig.~\ref{fig:time dependence}.II(b)). The strain-induced contrast increases following excitation and reaches a plateau after approximately 5-10 ns, remaining nearly constant up to the longest delay shown ($\sim50$ ns). The persistence of the birefringence signal indicates the presence of a long-lived, quasi-static strain component, consistent with thermoelastic deformation of the excited volume. 

Comparison of the two responses shows that both strain-induced birefringence and reduced magnetic contrast develop promptly following excitation. The magnetic contrast subsequently evolves toward reversal after approximately 1.5-2 ns, while the strain signal progressively increases toward a plateau. We note that the magnetic and structural dynamics shown here were measured at different pump wavelengths and pulse energies, and that the timescale of the magnetic response depends on the excitation strength (see Supplemental Material 5). Nevertheless, magnetic contrast reversal and a strain-sensitive response are observed over a broad range of pump wavelengths, suggesting that both magnetic and structural responses develop on a similar timescale. Because the measurements were performed under different excitation conditions, however, a direct quantitative comparison of their temporal evolution is not possible.

\begin{figure}
    \centering
    \includegraphics[width=0.8\linewidth]{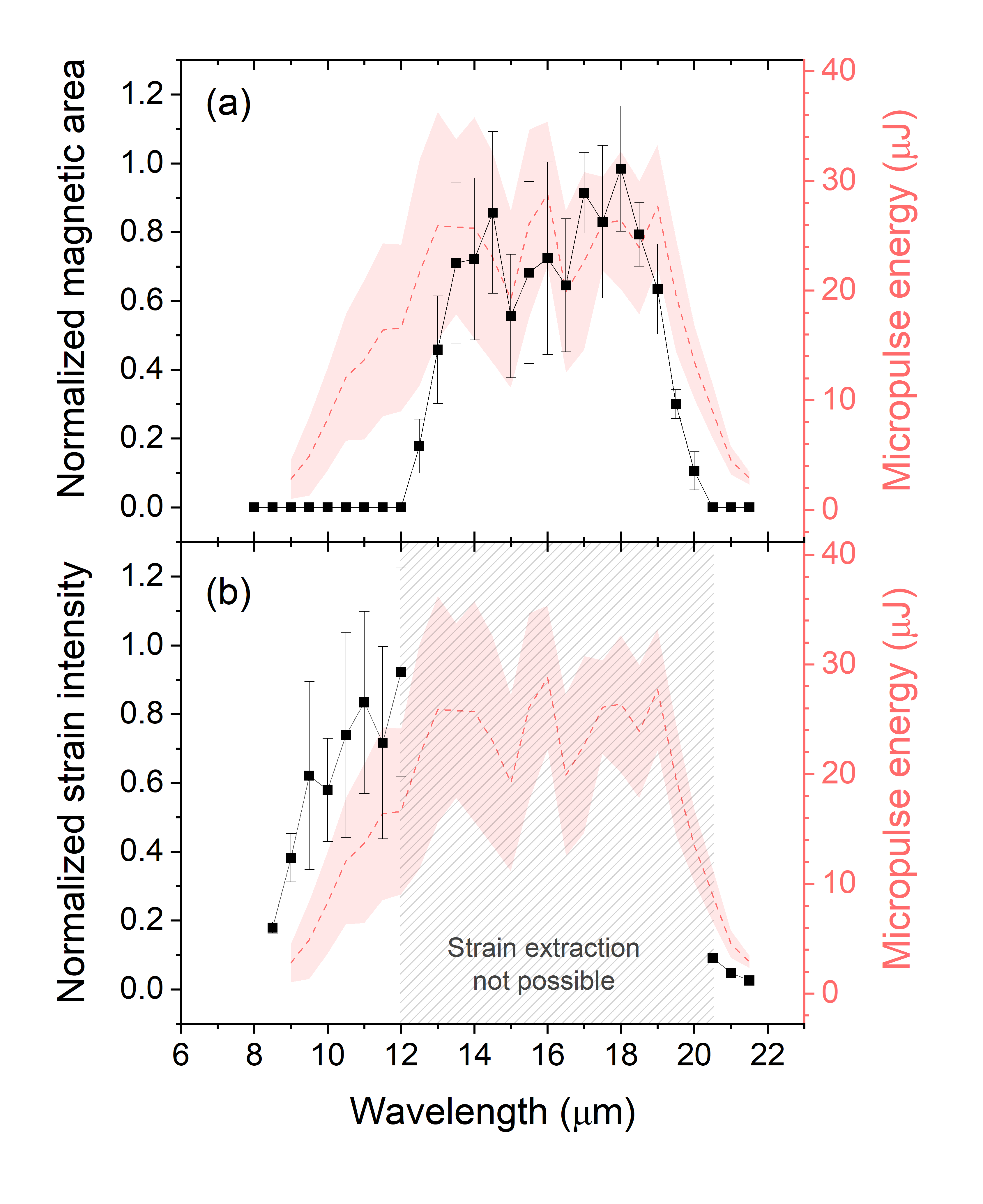}
    \caption{(a) Wavelength dependence of the magnetic area affected by the micropulse. (b) Wavelength dependence of the pixel intensity of the 4-lobe pattern associated to strain. In the shaded wavelength range 12-20 $\mu$m, magnetization switching overlaps the strain pattern, preventing the proper extraction of the strain intensity.}
    \label{fig:wavelength dependence}
\end{figure}

Further evidence linking the structural and magnetic responses is provided by their excitation-wavelength dependence. Figure~\ref{fig:wavelength dependence}(a-b) compares the magnetic and strain signals over the accessible spectral range from 8 to 22~$\mu$m (see Supplemental Material 6). Both responses are observed over a broad wavelength range and exhibit correlated dependences on excitation wavelength and pulse energy. This correspondence together with the micromagnetic simulations discussed below, supports a strain-mediated origin of the observed magnetization dynamics.\\

\begin{figure}[h!]
    \centering
    \includegraphics[width=1\linewidth]{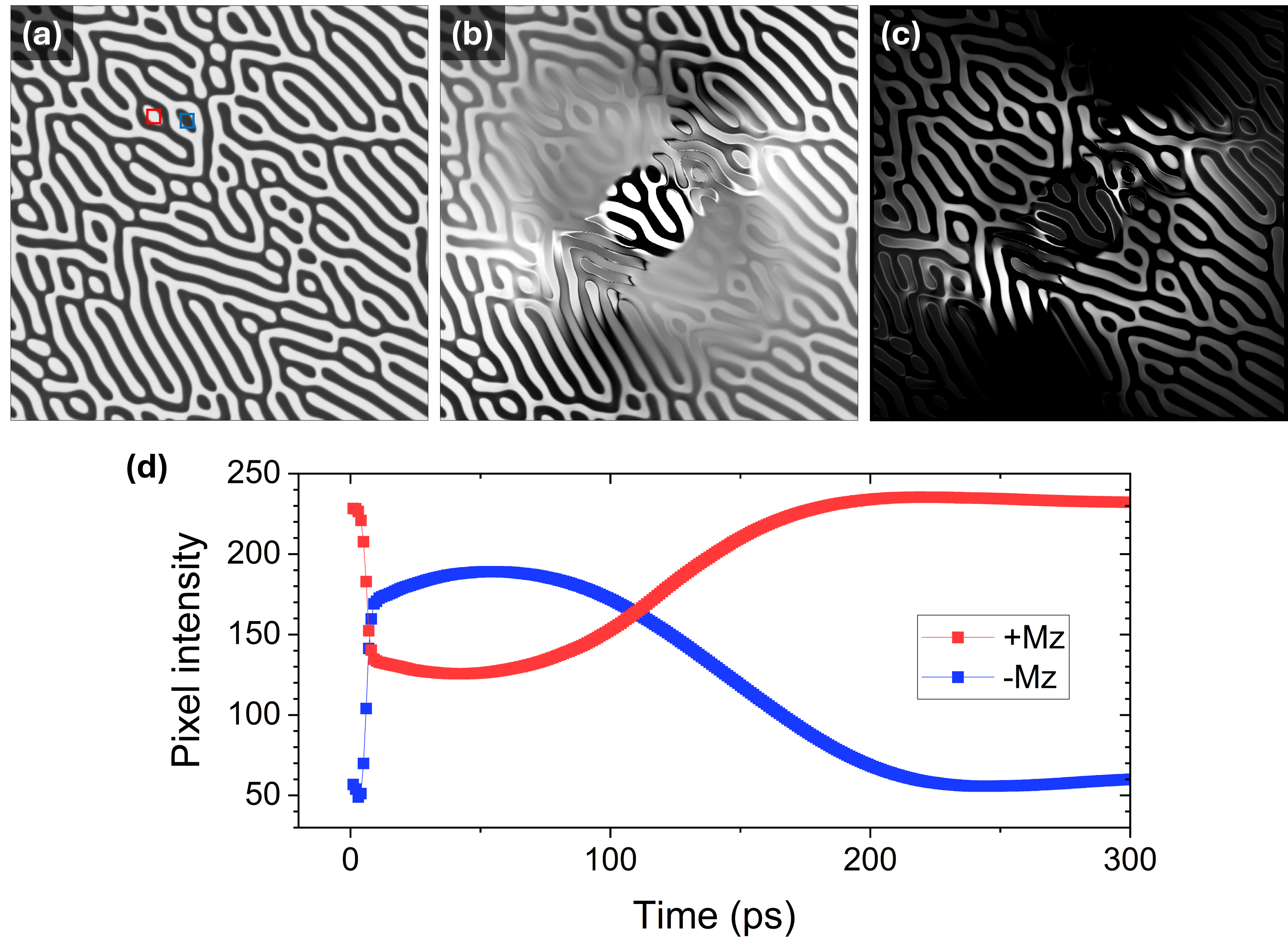}
    \caption{Simulated magnetic state (a) before, (b) 30 ps after the application of a 5-ps-long strain pulse and (c) the subtracted image. }
    \label{fig:simulation}
\end{figure}

To investigate the influence of transient strain on the magnetic domain configuration, we performed micromagnetic simulations using MuMax3 \cite{vansteenkiste2014design}. The simulated magnetic layer had dimensions of $10~\mu\mathrm{m}\times10~\mu\mathrm{m}\times1~\mu\mathrm{m}$ and was discretized into $10~\mathrm{nm}\times10~\mathrm{nm}\times1~\mu\mathrm{m}$ cells. The material parameters were an exchange stiffness of $A_{\mathrm{exc}}=3.7\times10^{-12}$~J/m \cite{klingler2014measurements}, cubic and uniaxial anisotropy constants of $K_c=-1.5\times10^3$~J/m$^3$ and $K_u=5\times10^2$~J/m$^3$ \cite{stupakiewicz2017ultrafast}, and magnetoelastic coupling coefficients of $b_1=3.48\times10^5$~J/m$^3$ and $b_2=6.96\times10^5$~J/m$^3$ \cite{smith1963magnetostriction}. The saturation magnetization and Gilbert damping parameter were set to $M_s=250$~kA/m and $\alpha=0.5$, respectively \cite{stupakiewicz2017ultrafast}. Transient strain was incorporated following Refs.~\cite{stupakiewicz2021ultrafast,gidding2023dynamic,damas2026photoinduced}, as detailed in Supplemental Material 3. The initial out-of-plane magnetization shown in Fig.~\ref{fig:simulation}(a) reproduces a labyrinthine domain state comparable to that observed experimentally. Thirty picoseconds after application of the strain pulse, a reversal of the out-of-plane magnetization component is obtained within individual domains (Fig.~\ref{fig:simulation}(b)), while the overall domain morphology remains largely preserved.

The corresponding difference image (Fig.~\ref{fig:simulation}(c)) reveals a spatially inhomogeneous response, with the strongest changes concentrated in two lobes and a comparatively weak response near the center of the excitation region, in contrast to the experimental observations. This discrepancy may arise from the simplified simulation geometry and the imposed two-dimensional strain distribution, which neglects the full three-dimensional character of strain propagation and its depth dependence. A more realistic treatment incorporating a three-dimensional strain field, as considered in Ref.~\cite{lourens2026ferroelastic}, may be required to reproduce the experimentally observed spatial profile. The present simulations are therefore intended to elucidate the local magnetic response to transient strain rather than quantitatively reproduce the spatial distribution of the measured dynamics. Despite this difference, the temporal evolution of the magnetization within representative domains (Fig.~\ref{fig:simulation}(d)) exhibits an oscillatory response characteristic of magnetization precession. The calculations therefore reproduce the principal qualitative features of the experiment: transient precessional reorientation of the magnetization within existing domains while the overall labyrinthine domain morphology remains largely preserved. The simulations consequently demonstrate that transient strain is sufficient to generate the type of magnetization dynamics observed experimentally. \\

Our experimental observations and micromagnetic simulations support a strain-mediated mechanism for the mid-infrared-induced magnetization dynamics in Co:YIG. A single 5-ps pulse generates a localized, long-lived strain-sensitive response and a propagating acoustic wave, accompanied by a transient reversal of the out-of-plane magnetic contrast on a nanosecond timescale. Remarkably, this large-amplitude magnetic response is reversible, while the labyrinthine domain morphology remains largely preserved. Micromagnetic simulations further show that transient strain can drive precessional magnetization reorientation within existing domains without substantial domain reorganization. These findings highlight single-pulse magnetoelastic excitation as a route to reversible, large-amplitude magnetization dynamics in multidomain magnetic structures at zero applied magnetic field. \\

\textit{Acknowledgements--} We thank the technical staff at FELIX for providing technical support. This publication is part of the project NL-ECO: Netherlands Initiative for Energy-Efficient Computing (with project number NWA. 1389.20.140) of the NWA research programme Research Along Routes by Consortia which is financed by the Dutch Research Council (NWO). H.D., C.S.D. and T.Z acknowledge support from the European Research Council ERC Grant Agreement No. 101115234 (HandShake), and A.K. acknowledges support from the European Research Council ERC Grant Agreement No. 101141740 (INTERPHON).


\begin{thebibliography}{47}%
\makeatletter
\providecommand \@ifxundefined [1]{%
 \@ifx{#1\undefined}
}%
\providecommand \@ifnum [1]{%
 \ifnum #1\expandafter \@firstoftwo
 \else \expandafter \@secondoftwo
 \fi
}%
\providecommand \@ifx [1]{%
 \ifx #1\expandafter \@firstoftwo
 \else \expandafter \@secondoftwo
 \fi
}%
\providecommand \natexlab [1]{#1}%
\providecommand \enquote  [1]{``#1''}%
\providecommand \bibnamefont  [1]{#1}%
\providecommand \bibfnamefont [1]{#1}%
\providecommand \citenamefont [1]{#1}%
\providecommand \href@noop [0]{\@secondoftwo}%
\providecommand \href [0]{\begingroup \@sanitize@url \@href}%
\providecommand \@href[1]{\@@startlink{#1}\@@href}%
\providecommand \@@href[1]{\endgroup#1\@@endlink}%
\providecommand \@sanitize@url [0]{\catcode `\\12\catcode `\$12\catcode `\&12\catcode `\#12\catcode `\^12\catcode `\_12\catcode `\%12\relax}%
\providecommand \@@startlink[1]{}%
\providecommand \@@endlink[0]{}%
\providecommand \url  [0]{\begingroup\@sanitize@url \@url }%
\providecommand \@url [1]{\endgroup\@href {#1}{\urlprefix }}%
\providecommand \urlprefix  [0]{URL }%
\providecommand \Eprint [0]{\href }%
\providecommand \doibase [0]{https://doi.org/}%
\providecommand \selectlanguage [0]{\@gobble}%
\providecommand \bibinfo  [0]{\@secondoftwo}%
\providecommand \bibfield  [0]{\@secondoftwo}%
\providecommand \translation [1]{[#1]}%
\providecommand \BibitemOpen [0]{}%
\providecommand \bibitemStop [0]{}%
\providecommand \bibitemNoStop [0]{.\EOS\space}%
\providecommand \EOS [0]{\spacefactor3000\relax}%
\providecommand \BibitemShut  [1]{\csname bibitem#1\endcsname}%
\let\auto@bib@innerbib\@empty
\bibitem [{\citenamefont {Back}\ \emph {et~al.}(1999)\citenamefont {Back}, \citenamefont {Allenspach}, \citenamefont {Weber}, \citenamefont {Parkin}, \citenamefont {Weller}, \citenamefont {Garwin},\ and\ \citenamefont {Siegmann}}]{back1999minimum}%
  \BibitemOpen
  \bibfield  {author} {\bibinfo {author} {\bibfnamefont {C.}~\bibnamefont {Back}}, \bibinfo {author} {\bibfnamefont {R.}~\bibnamefont {Allenspach}}, \bibinfo {author} {\bibfnamefont {W.}~\bibnamefont {Weber}}, \bibinfo {author} {\bibfnamefont {S.}~\bibnamefont {Parkin}}, \bibinfo {author} {\bibfnamefont {D.}~\bibnamefont {Weller}}, \bibinfo {author} {\bibfnamefont {E.~L.}\ \bibnamefont {Garwin}},\ and\ \bibinfo {author} {\bibfnamefont {H.}~\bibnamefont {Siegmann}},\ }\bibfield  {title} {\bibinfo {title} {Minimum field strength in precessional magnetization reversal},\ }\href@noop {} {\bibfield  {journal} {\bibinfo  {journal} {Science}\ }\textbf {\bibinfo {volume} {285}},\ \bibinfo {pages} {864} (\bibinfo {year} {1999})}\BibitemShut {NoStop}%
\bibitem [{\citenamefont {Gerrits}\ \emph {et~al.}(2002)\citenamefont {Gerrits}, \citenamefont {Van Den~Berg}, \citenamefont {Hohlfeld}, \citenamefont {B{\"a}r},\ and\ \citenamefont {Rasing}}]{gerrits2002ultrafast}%
  \BibitemOpen
  \bibfield  {author} {\bibinfo {author} {\bibfnamefont {T.}~\bibnamefont {Gerrits}}, \bibinfo {author} {\bibfnamefont {H.}~\bibnamefont {Van Den~Berg}}, \bibinfo {author} {\bibfnamefont {J.}~\bibnamefont {Hohlfeld}}, \bibinfo {author} {\bibfnamefont {L.}~\bibnamefont {B{\"a}r}},\ and\ \bibinfo {author} {\bibfnamefont {T.}~\bibnamefont {Rasing}},\ }\bibfield  {title} {\bibinfo {title} {Ultrafast precessional magnetization reversal by picosecond magnetic field pulse shaping},\ }\href@noop {} {\bibfield  {journal} {\bibinfo  {journal} {Nature}\ }\textbf {\bibinfo {volume} {418}},\ \bibinfo {pages} {509} (\bibinfo {year} {2002})}\BibitemShut {NoStop}%
\bibitem [{\citenamefont {Kimel}\ \emph {et~al.}(2005)\citenamefont {Kimel}, \citenamefont {Kirilyuk}, \citenamefont {Usachev}, \citenamefont {Pisarev}, \citenamefont {Balbashov},\ and\ \citenamefont {Rasing}}]{kimel2005ultrafast}%
  \BibitemOpen
  \bibfield  {author} {\bibinfo {author} {\bibfnamefont {A.}~\bibnamefont {Kimel}}, \bibinfo {author} {\bibfnamefont {A.}~\bibnamefont {Kirilyuk}}, \bibinfo {author} {\bibfnamefont {P.}~\bibnamefont {Usachev}}, \bibinfo {author} {\bibfnamefont {R.}~\bibnamefont {Pisarev}}, \bibinfo {author} {\bibfnamefont {A.}~\bibnamefont {Balbashov}},\ and\ \bibinfo {author} {\bibfnamefont {T.}~\bibnamefont {Rasing}},\ }\bibfield  {title} {\bibinfo {title} {Ultrafast non-thermal control of magnetization by instantaneous photomagnetic pulses},\ }\href@noop {} {\bibfield  {journal} {\bibinfo  {journal} {Nature}\ }\textbf {\bibinfo {volume} {435}},\ \bibinfo {pages} {655} (\bibinfo {year} {2005})}\BibitemShut {NoStop}%
\bibitem [{\citenamefont {Van~Kampen}\ \emph {et~al.}(2002)\citenamefont {Van~Kampen}, \citenamefont {Jozsa}, \citenamefont {Kohlhepp}, \citenamefont {LeClair}, \citenamefont {Lagae}, \citenamefont {De~Jonge},\ and\ \citenamefont {Koopmans}}]{van2002all}%
  \BibitemOpen
  \bibfield  {author} {\bibinfo {author} {\bibfnamefont {M.}~\bibnamefont {Van~Kampen}}, \bibinfo {author} {\bibfnamefont {C.}~\bibnamefont {Jozsa}}, \bibinfo {author} {\bibfnamefont {J.}~\bibnamefont {Kohlhepp}}, \bibinfo {author} {\bibfnamefont {P.}~\bibnamefont {LeClair}}, \bibinfo {author} {\bibfnamefont {L.}~\bibnamefont {Lagae}}, \bibinfo {author} {\bibfnamefont {W.}~\bibnamefont {De~Jonge}},\ and\ \bibinfo {author} {\bibfnamefont {B.}~\bibnamefont {Koopmans}},\ }\bibfield  {title} {\bibinfo {title} {All-optical probe of coherent spin waves},\ }\href@noop {} {\bibfield  {journal} {\bibinfo  {journal} {Physical review letters}\ }\textbf {\bibinfo {volume} {88}},\ \bibinfo {pages} {227201} (\bibinfo {year} {2002})}\BibitemShut {NoStop}%
\bibitem [{\citenamefont {Scherbakov}\ \emph {et~al.}(2010)\citenamefont {Scherbakov}, \citenamefont {Salasyuk}, \citenamefont {Akimov}, \citenamefont {Liu}, \citenamefont {Bombeck}, \citenamefont {Br{\"u}ggemann}, \citenamefont {Yakovlev}, \citenamefont {Sapega}, \citenamefont {Furdyna},\ and\ \citenamefont {Bayer}}]{scherbakov2010coherent}%
  \BibitemOpen
  \bibfield  {author} {\bibinfo {author} {\bibfnamefont {A.}~\bibnamefont {Scherbakov}}, \bibinfo {author} {\bibfnamefont {A.}~\bibnamefont {Salasyuk}}, \bibinfo {author} {\bibfnamefont {A.}~\bibnamefont {Akimov}}, \bibinfo {author} {\bibfnamefont {X.}~\bibnamefont {Liu}}, \bibinfo {author} {\bibfnamefont {M.}~\bibnamefont {Bombeck}}, \bibinfo {author} {\bibfnamefont {C.}~\bibnamefont {Br{\"u}ggemann}}, \bibinfo {author} {\bibfnamefont {D.}~\bibnamefont {Yakovlev}}, \bibinfo {author} {\bibfnamefont {V.}~\bibnamefont {Sapega}}, \bibinfo {author} {\bibfnamefont {J.}~\bibnamefont {Furdyna}},\ and\ \bibinfo {author} {\bibfnamefont {M.}~\bibnamefont {Bayer}},\ }\bibfield  {title} {\bibinfo {title} {Coherent magnetization precession in ferromagnetic (ga, mn)as induced by picosecond acoustic pulses},\ }\href@noop {} {\bibfield  {journal} {\bibinfo  {journal} {Physical review letters}\ }\textbf {\bibinfo {volume} {105}},\ \bibinfo {pages} {117204} (\bibinfo {year} {2010})}\BibitemShut {NoStop}%
\bibitem [{\citenamefont {Linnik}\ \emph {et~al.}(2011)\citenamefont {Linnik}, \citenamefont {Scherbakov}, \citenamefont {Yakovlev}, \citenamefont {Liu}, \citenamefont {Furdyna},\ and\ \citenamefont {Bayer}}]{linnik2011theory}%
  \BibitemOpen
  \bibfield  {author} {\bibinfo {author} {\bibfnamefont {T.}~\bibnamefont {Linnik}}, \bibinfo {author} {\bibfnamefont {A.}~\bibnamefont {Scherbakov}}, \bibinfo {author} {\bibfnamefont {D.}~\bibnamefont {Yakovlev}}, \bibinfo {author} {\bibfnamefont {X.}~\bibnamefont {Liu}}, \bibinfo {author} {\bibfnamefont {J.}~\bibnamefont {Furdyna}},\ and\ \bibinfo {author} {\bibfnamefont {M.}~\bibnamefont {Bayer}},\ }\bibfield  {title} {\bibinfo {title} {Theory of magnetization precession induced by a picosecond strain pulse in ferromagnetic semiconductor (ga, mn) as},\ }\href@noop {} {\bibfield  {journal} {\bibinfo  {journal} {Physical Review B—Condensed Matter and Materials Physics}\ }\textbf {\bibinfo {volume} {84}},\ \bibinfo {pages} {214432} (\bibinfo {year} {2011})}\BibitemShut {NoStop}%
\bibitem [{\citenamefont {Kim}\ \emph {et~al.}(2012)\citenamefont {Kim}, \citenamefont {Vomir},\ and\ \citenamefont {Bigot}}]{kim2012ultrafast}%
  \BibitemOpen
  \bibfield  {author} {\bibinfo {author} {\bibfnamefont {J.-W.}\ \bibnamefont {Kim}}, \bibinfo {author} {\bibfnamefont {M.}~\bibnamefont {Vomir}},\ and\ \bibinfo {author} {\bibfnamefont {J.-Y.}\ \bibnamefont {Bigot}},\ }\bibfield  {title} {\bibinfo {title} {Ultrafast magnetoacoustics in nickel films},\ }\href@noop {} {\bibfield  {journal} {\bibinfo  {journal} {Physical review letters}\ }\textbf {\bibinfo {volume} {109}},\ \bibinfo {pages} {166601} (\bibinfo {year} {2012})}\BibitemShut {NoStop}%
\bibitem [{\citenamefont {Kovalenko}\ \emph {et~al.}(2013)\citenamefont {Kovalenko}, \citenamefont {Pezeril},\ and\ \citenamefont {Temnov}}]{kovalenko2013new}%
  \BibitemOpen
  \bibfield  {author} {\bibinfo {author} {\bibfnamefont {O.}~\bibnamefont {Kovalenko}}, \bibinfo {author} {\bibfnamefont {T.}~\bibnamefont {Pezeril}},\ and\ \bibinfo {author} {\bibfnamefont {V.~V.}\ \bibnamefont {Temnov}},\ }\bibfield  {title} {\bibinfo {title} {New concept for magnetization switching by ultrafast acoustic pulses},\ }\href@noop {} {\bibfield  {journal} {\bibinfo  {journal} {Physical review letters}\ }\textbf {\bibinfo {volume} {110}},\ \bibinfo {pages} {266602} (\bibinfo {year} {2013})}\BibitemShut {NoStop}%
\bibitem [{\citenamefont {Thevenard}\ \emph {et~al.}(2016)\citenamefont {Thevenard}, \citenamefont {Camara}, \citenamefont {Majrab}, \citenamefont {Bernard}, \citenamefont {Rovillain}, \citenamefont {Lema{\^\i}tre}, \citenamefont {Gourdon},\ and\ \citenamefont {Duquesne}}]{thevenard2016precessional}%
  \BibitemOpen
  \bibfield  {author} {\bibinfo {author} {\bibfnamefont {L.}~\bibnamefont {Thevenard}}, \bibinfo {author} {\bibfnamefont {I.~S.}\ \bibnamefont {Camara}}, \bibinfo {author} {\bibfnamefont {S.}~\bibnamefont {Majrab}}, \bibinfo {author} {\bibfnamefont {M.}~\bibnamefont {Bernard}}, \bibinfo {author} {\bibfnamefont {P.}~\bibnamefont {Rovillain}}, \bibinfo {author} {\bibfnamefont {A.}~\bibnamefont {Lema{\^\i}tre}}, \bibinfo {author} {\bibfnamefont {C.}~\bibnamefont {Gourdon}},\ and\ \bibinfo {author} {\bibfnamefont {J.-Y.}\ \bibnamefont {Duquesne}},\ }\bibfield  {title} {\bibinfo {title} {Precessional magnetization switching by a surface acoustic wave},\ }\href@noop {} {\bibfield  {journal} {\bibinfo  {journal} {Physical Review B}\ }\textbf {\bibinfo {volume} {93}},\ \bibinfo {pages} {134430} (\bibinfo {year} {2016})}\BibitemShut {NoStop}%
\bibitem [{\citenamefont {Zeiger}\ \emph {et~al.}(1992)\citenamefont {Zeiger}, \citenamefont {Vidal}, \citenamefont {Cheng}, \citenamefont {Ippen}, \citenamefont {Dresselhaus},\ and\ \citenamefont {Dresselhaus}}]{zeiger1992theory}%
  \BibitemOpen
  \bibfield  {author} {\bibinfo {author} {\bibfnamefont {H.}~\bibnamefont {Zeiger}}, \bibinfo {author} {\bibfnamefont {J.}~\bibnamefont {Vidal}}, \bibinfo {author} {\bibfnamefont {T.}~\bibnamefont {Cheng}}, \bibinfo {author} {\bibfnamefont {E.}~\bibnamefont {Ippen}}, \bibinfo {author} {\bibfnamefont {G.}~\bibnamefont {Dresselhaus}},\ and\ \bibinfo {author} {\bibfnamefont {M.}~\bibnamefont {Dresselhaus}},\ }\bibfield  {title} {\bibinfo {title} {Theory for displacive excitation of coherent phonons},\ }\href@noop {} {\bibfield  {journal} {\bibinfo  {journal} {Physical Review B}\ }\textbf {\bibinfo {volume} {45}},\ \bibinfo {pages} {768} (\bibinfo {year} {1992})}\BibitemShut {NoStop}%
\bibitem [{\citenamefont {Merlin}(1997)}]{merlin1997generating}%
  \BibitemOpen
  \bibfield  {author} {\bibinfo {author} {\bibfnamefont {R.}~\bibnamefont {Merlin}},\ }\bibfield  {title} {\bibinfo {title} {Generating coherent thz phonons with light pulses},\ }\href@noop {} {\bibfield  {journal} {\bibinfo  {journal} {Solid state communications}\ }\textbf {\bibinfo {volume} {102}},\ \bibinfo {pages} {207} (\bibinfo {year} {1997})}\BibitemShut {NoStop}%
\bibitem [{\citenamefont {Thomsen}\ \emph {et~al.}(1986)\citenamefont {Thomsen}, \citenamefont {Grahn}, \citenamefont {Maris},\ and\ \citenamefont {Tauc}}]{thomsen1986surface}%
  \BibitemOpen
  \bibfield  {author} {\bibinfo {author} {\bibfnamefont {C.}~\bibnamefont {Thomsen}}, \bibinfo {author} {\bibfnamefont {H.~T.}\ \bibnamefont {Grahn}}, \bibinfo {author} {\bibfnamefont {H.~J.}\ \bibnamefont {Maris}},\ and\ \bibinfo {author} {\bibfnamefont {J.}~\bibnamefont {Tauc}},\ }\bibfield  {title} {\bibinfo {title} {Surface generation and detection of phonons by picosecond light pulses},\ }\href@noop {} {\bibfield  {journal} {\bibinfo  {journal} {Physical Review B}\ }\textbf {\bibinfo {volume} {34}},\ \bibinfo {pages} {4129} (\bibinfo {year} {1986})}\BibitemShut {NoStop}%
\bibitem [{\citenamefont {Ruello}\ and\ \citenamefont {Gusev}(2015)}]{ruello2015physical}%
  \BibitemOpen
  \bibfield  {author} {\bibinfo {author} {\bibfnamefont {P.}~\bibnamefont {Ruello}}\ and\ \bibinfo {author} {\bibfnamefont {V.~E.}\ \bibnamefont {Gusev}},\ }\bibfield  {title} {\bibinfo {title} {Physical mechanisms of coherent acoustic phonons generation by ultrafast laser action},\ }\href@noop {} {\bibfield  {journal} {\bibinfo  {journal} {Ultrasonics}\ }\textbf {\bibinfo {volume} {56}},\ \bibinfo {pages} {21} (\bibinfo {year} {2015})}\BibitemShut {NoStop}%
\bibitem [{\citenamefont {Lejman}\ \emph {et~al.}(2014)\citenamefont {Lejman}, \citenamefont {Vaudel}, \citenamefont {Infante}, \citenamefont {Gemeiner}, \citenamefont {Gusev}, \citenamefont {Dkhil},\ and\ \citenamefont {Ruello}}]{lejman2014giant}%
  \BibitemOpen
  \bibfield  {author} {\bibinfo {author} {\bibfnamefont {M.}~\bibnamefont {Lejman}}, \bibinfo {author} {\bibfnamefont {G.}~\bibnamefont {Vaudel}}, \bibinfo {author} {\bibfnamefont {I.~C.}\ \bibnamefont {Infante}}, \bibinfo {author} {\bibfnamefont {P.}~\bibnamefont {Gemeiner}}, \bibinfo {author} {\bibfnamefont {V.~E.}\ \bibnamefont {Gusev}}, \bibinfo {author} {\bibfnamefont {B.}~\bibnamefont {Dkhil}},\ and\ \bibinfo {author} {\bibfnamefont {P.}~\bibnamefont {Ruello}},\ }\bibfield  {title} {\bibinfo {title} {Giant ultrafast photo-induced shear strain in ferroelectric bifeo3},\ }\href@noop {} {\bibfield  {journal} {\bibinfo  {journal} {Nature communications}\ }\textbf {\bibinfo {volume} {5}},\ \bibinfo {pages} {4301} (\bibinfo {year} {2014})}\BibitemShut {NoStop}%
\bibitem [{\citenamefont {Shin}\ \emph {et~al.}(2022)\citenamefont {Shin}, \citenamefont {Vomir}, \citenamefont {Kim}, \citenamefont {Van}, \citenamefont {Jeong},\ and\ \citenamefont {Kim}}]{shin2022quasi}%
  \BibitemOpen
  \bibfield  {author} {\bibinfo {author} {\bibfnamefont {Y.}~\bibnamefont {Shin}}, \bibinfo {author} {\bibfnamefont {M.}~\bibnamefont {Vomir}}, \bibinfo {author} {\bibfnamefont {D.-H.}\ \bibnamefont {Kim}}, \bibinfo {author} {\bibfnamefont {P.~C.}\ \bibnamefont {Van}}, \bibinfo {author} {\bibfnamefont {J.-R.}\ \bibnamefont {Jeong}},\ and\ \bibinfo {author} {\bibfnamefont {J.-W.}\ \bibnamefont {Kim}},\ }\bibfield  {title} {\bibinfo {title} {Quasi-static strain governing ultrafast spin dynamics},\ }\href@noop {} {\bibfield  {journal} {\bibinfo  {journal} {Communications Physics}\ }\textbf {\bibinfo {volume} {5}},\ \bibinfo {pages} {56} (\bibinfo {year} {2022})}\BibitemShut {NoStop}%
\bibitem [{\citenamefont {Walz}\ \emph {et~al.}(2025)\citenamefont {Walz}, \citenamefont {Weber}, \citenamefont {Zeuschner}, \citenamefont {Dumesnil}, \citenamefont {von Reppert},\ and\ \citenamefont {Bargheer}}]{walz2025large}%
  \BibitemOpen
  \bibfield  {author} {\bibinfo {author} {\bibfnamefont {C.}~\bibnamefont {Walz}}, \bibinfo {author} {\bibfnamefont {F.-C.}\ \bibnamefont {Weber}}, \bibinfo {author} {\bibfnamefont {S.-P.}\ \bibnamefont {Zeuschner}}, \bibinfo {author} {\bibfnamefont {K.}~\bibnamefont {Dumesnil}}, \bibinfo {author} {\bibfnamefont {A.}~\bibnamefont {von Reppert}},\ and\ \bibinfo {author} {\bibfnamefont {M.}~\bibnamefont {Bargheer}},\ }\bibfield  {title} {\bibinfo {title} {Large strain contribution to the laser-driven magnetization response of magnetostrictive tbfe2},\ }\href@noop {} {\bibfield  {journal} {\bibinfo  {journal} {Applied Physics Letters}\ }\textbf {\bibinfo {volume} {127}} (\bibinfo {year} {2025})}\BibitemShut {NoStop}%
\bibitem [{\citenamefont {Nicoletti}\ and\ \citenamefont {Cavalleri}(2016)}]{nicoletti2016nonlinear}%
  \BibitemOpen
  \bibfield  {author} {\bibinfo {author} {\bibfnamefont {D.}~\bibnamefont {Nicoletti}}\ and\ \bibinfo {author} {\bibfnamefont {A.}~\bibnamefont {Cavalleri}},\ }\bibfield  {title} {\bibinfo {title} {Nonlinear light--matter interaction at terahertz frequencies},\ }\href@noop {} {\bibfield  {journal} {\bibinfo  {journal} {Advances in Optics and Photonics}\ }\textbf {\bibinfo {volume} {8}},\ \bibinfo {pages} {401} (\bibinfo {year} {2016})}\BibitemShut {NoStop}%
\bibitem [{\citenamefont {Subedi}(2021)}]{subedi2021light}%
  \BibitemOpen
  \bibfield  {author} {\bibinfo {author} {\bibfnamefont {A.}~\bibnamefont {Subedi}},\ }\bibfield  {title} {\bibinfo {title} {Light-control of materials via nonlinear phononics},\ }\href@noop {} {\bibfield  {journal} {\bibinfo  {journal} {Comptes Rendus. Physique}\ }\textbf {\bibinfo {volume} {22}},\ \bibinfo {pages} {161} (\bibinfo {year} {2021})}\BibitemShut {NoStop}%
\bibitem [{\citenamefont {F{\"o}rst}\ \emph {et~al.}(2011)\citenamefont {F{\"o}rst}, \citenamefont {Manzoni}, \citenamefont {Kaiser}, \citenamefont {Tomioka}, \citenamefont {Tokura}, \citenamefont {Merlin},\ and\ \citenamefont {Cavalleri}}]{forst2011nonlinear}%
  \BibitemOpen
  \bibfield  {author} {\bibinfo {author} {\bibfnamefont {M.}~\bibnamefont {F{\"o}rst}}, \bibinfo {author} {\bibfnamefont {C.}~\bibnamefont {Manzoni}}, \bibinfo {author} {\bibfnamefont {S.}~\bibnamefont {Kaiser}}, \bibinfo {author} {\bibfnamefont {Y.}~\bibnamefont {Tomioka}}, \bibinfo {author} {\bibfnamefont {Y.}~\bibnamefont {Tokura}}, \bibinfo {author} {\bibfnamefont {R.}~\bibnamefont {Merlin}},\ and\ \bibinfo {author} {\bibfnamefont {A.}~\bibnamefont {Cavalleri}},\ }\bibfield  {title} {\bibinfo {title} {Nonlinear phononics as an ultrafast route to lattice control},\ }\href@noop {} {\bibfield  {journal} {\bibinfo  {journal} {Nature Physics}\ }\textbf {\bibinfo {volume} {7}},\ \bibinfo {pages} {854} (\bibinfo {year} {2011})}\BibitemShut {NoStop}%
\bibitem [{\citenamefont {Mankowsky}\ \emph {et~al.}(2016)\citenamefont {Mankowsky}, \citenamefont {F{\"o}rst},\ and\ \citenamefont {Cavalleri}}]{mankowsky2016non}%
  \BibitemOpen
  \bibfield  {author} {\bibinfo {author} {\bibfnamefont {R.}~\bibnamefont {Mankowsky}}, \bibinfo {author} {\bibfnamefont {M.}~\bibnamefont {F{\"o}rst}},\ and\ \bibinfo {author} {\bibfnamefont {A.}~\bibnamefont {Cavalleri}},\ }\bibfield  {title} {\bibinfo {title} {Non-equilibrium control of complex solids by nonlinear phononics},\ }\href@noop {} {\bibfield  {journal} {\bibinfo  {journal} {Reports on Progress in Physics}\ }\textbf {\bibinfo {volume} {79}},\ \bibinfo {pages} {064503} (\bibinfo {year} {2016})}\BibitemShut {NoStop}%
\bibitem [{\citenamefont {Disa}\ \emph {et~al.}(2021)\citenamefont {Disa}, \citenamefont {Nova},\ and\ \citenamefont {Cavalleri}}]{disa2021engineering}%
  \BibitemOpen
  \bibfield  {author} {\bibinfo {author} {\bibfnamefont {A.~S.}\ \bibnamefont {Disa}}, \bibinfo {author} {\bibfnamefont {T.~F.}\ \bibnamefont {Nova}},\ and\ \bibinfo {author} {\bibfnamefont {A.}~\bibnamefont {Cavalleri}},\ }\bibfield  {title} {\bibinfo {title} {Engineering crystal structures with light},\ }\href@noop {} {\bibfield  {journal} {\bibinfo  {journal} {Nature Physics}\ }\textbf {\bibinfo {volume} {17}},\ \bibinfo {pages} {1087} (\bibinfo {year} {2021})}\BibitemShut {NoStop}%
\bibitem [{\citenamefont {Davies}\ \emph {et~al.}(2024)\citenamefont {Davies}, \citenamefont {Fennema}, \citenamefont {Tsukamoto}, \citenamefont {Razdolski}, \citenamefont {Kimel},\ and\ \citenamefont {Kirilyuk}}]{davies2024phononic}%
  \BibitemOpen
  \bibfield  {author} {\bibinfo {author} {\bibfnamefont {C.~S.}\ \bibnamefont {Davies}}, \bibinfo {author} {\bibfnamefont {F.}~\bibnamefont {Fennema}}, \bibinfo {author} {\bibfnamefont {A.}~\bibnamefont {Tsukamoto}}, \bibinfo {author} {\bibfnamefont {I.}~\bibnamefont {Razdolski}}, \bibinfo {author} {\bibfnamefont {A.~V.}\ \bibnamefont {Kimel}},\ and\ \bibinfo {author} {\bibfnamefont {A.}~\bibnamefont {Kirilyuk}},\ }\bibfield  {title} {\bibinfo {title} {Phononic switching of magnetization by the ultrafast barnett effect},\ }\href@noop {} {\bibfield  {journal} {\bibinfo  {journal} {Nature}\ }\textbf {\bibinfo {volume} {628}},\ \bibinfo {pages} {540} (\bibinfo {year} {2024})}\BibitemShut {NoStop}%
\bibitem [{\citenamefont {Fennema}\ \emph {et~al.}(2026)\citenamefont {Fennema}, \citenamefont {Damas}, \citenamefont {Yoshikawa}, \citenamefont {Tsukamoto}, \citenamefont {Davies},\ and\ \citenamefont {Kirilyuk}}]{fennema2026generalized}%
  \BibitemOpen
  \bibfield  {author} {\bibinfo {author} {\bibfnamefont {F.}~\bibnamefont {Fennema}}, \bibinfo {author} {\bibfnamefont {H.}~\bibnamefont {Damas}}, \bibinfo {author} {\bibfnamefont {H.}~\bibnamefont {Yoshikawa}}, \bibinfo {author} {\bibfnamefont {A.}~\bibnamefont {Tsukamoto}}, \bibinfo {author} {\bibfnamefont {C.}~\bibnamefont {Davies}},\ and\ \bibinfo {author} {\bibfnamefont {A.}~\bibnamefont {Kirilyuk}},\ }\bibfield  {title} {\bibinfo {title} {Generalized helicity-dependent magnetization switching via substrate phonons},\ }\href@noop {} {\bibfield  {journal} {\bibinfo  {journal} {arXiv preprint arXiv:2609.05064}\ } (\bibinfo {year} {2026})}\BibitemShut {NoStop}%
\bibitem [{\citenamefont {Damas}\ \emph {et~al.}(2026)\citenamefont {Damas}, \citenamefont {Davies}, \citenamefont {Vetoshko}, \citenamefont {Belotelov}, \citenamefont {Stupakiewicz},\ and\ \citenamefont {Kirilyuk}}]{damas2026photoinduced}%
  \BibitemOpen
  \bibfield  {author} {\bibinfo {author} {\bibfnamefont {H.}~\bibnamefont {Damas}}, \bibinfo {author} {\bibfnamefont {C.~S.}\ \bibnamefont {Davies}}, \bibinfo {author} {\bibfnamefont {P.~M.}\ \bibnamefont {Vetoshko}}, \bibinfo {author} {\bibfnamefont {V.~I.}\ \bibnamefont {Belotelov}}, \bibinfo {author} {\bibfnamefont {A.}~\bibnamefont {Stupakiewicz}},\ and\ \bibinfo {author} {\bibfnamefont {A.}~\bibnamefont {Kirilyuk}},\ }\bibfield  {title} {\bibinfo {title} {Photoinduced switching of magnetization in the epsilon-near-zero regime},\ }\href@noop {} {\bibfield  {journal} {\bibinfo  {journal} {Physical Review Letters}\ }\textbf {\bibinfo {volume} {137}},\ \bibinfo {pages} {036704} (\bibinfo {year} {2026})}\BibitemShut {NoStop}%
\bibitem [{\citenamefont {Kwaaitaal}\ \emph {et~al.}(2024{\natexlab{a}})\citenamefont {Kwaaitaal}, \citenamefont {Lourens}, \citenamefont {Davies},\ and\ \citenamefont {Kirilyuk}}]{kwaaitaal2024epsilon}%
  \BibitemOpen
  \bibfield  {author} {\bibinfo {author} {\bibfnamefont {M.}~\bibnamefont {Kwaaitaal}}, \bibinfo {author} {\bibfnamefont {D.~G.}\ \bibnamefont {Lourens}}, \bibinfo {author} {\bibfnamefont {C.~S.}\ \bibnamefont {Davies}},\ and\ \bibinfo {author} {\bibfnamefont {A.}~\bibnamefont {Kirilyuk}},\ }\bibfield  {title} {\bibinfo {title} {Epsilon-near-zero regime enables permanent ultrafast all-optical reversal of ferroelectric polarization},\ }\href@noop {} {\bibfield  {journal} {\bibinfo  {journal} {Nature Photonics}\ ,\ \bibinfo {pages} {1}} (\bibinfo {year} {2024}{\natexlab{a}})}\BibitemShut {NoStop}%
\bibitem [{\citenamefont {Kwaaitaal}\ \emph {et~al.}(2024{\natexlab{b}})\citenamefont {Kwaaitaal}, \citenamefont {Lourens}, \citenamefont {Davies},\ and\ \citenamefont {Kirilyuk}}]{kwaaitaal2024disentangling}%
  \BibitemOpen
  \bibfield  {author} {\bibinfo {author} {\bibfnamefont {M.}~\bibnamefont {Kwaaitaal}}, \bibinfo {author} {\bibfnamefont {D.~G.}\ \bibnamefont {Lourens}}, \bibinfo {author} {\bibfnamefont {C.~S.}\ \bibnamefont {Davies}},\ and\ \bibinfo {author} {\bibfnamefont {A.}~\bibnamefont {Kirilyuk}},\ }\bibfield  {title} {\bibinfo {title} {Disentangling thermal birefringence and strain in the all-optical switching of ferroelectric polarization},\ }\href@noop {} {\bibfield  {journal} {\bibinfo  {journal} {Scientific Reports}\ }\textbf {\bibinfo {volume} {14}},\ \bibinfo {pages} {24956} (\bibinfo {year} {2024}{\natexlab{b}})}\BibitemShut {NoStop}%
\bibitem [{\citenamefont {Afanasiev}\ \emph {et~al.}(2021{\natexlab{a}})\citenamefont {Afanasiev}, \citenamefont {Hortensius}, \citenamefont {Ivanov}, \citenamefont {Sasani}, \citenamefont {Bousquet}, \citenamefont {Blanter}, \citenamefont {Mikhaylovskiy}, \citenamefont {Kimel},\ and\ \citenamefont {Caviglia}}]{afanasiev2021ultrafast}%
  \BibitemOpen
  \bibfield  {author} {\bibinfo {author} {\bibfnamefont {D.}~\bibnamefont {Afanasiev}}, \bibinfo {author} {\bibfnamefont {J.}~\bibnamefont {Hortensius}}, \bibinfo {author} {\bibfnamefont {B.}~\bibnamefont {Ivanov}}, \bibinfo {author} {\bibfnamefont {A.}~\bibnamefont {Sasani}}, \bibinfo {author} {\bibfnamefont {E.}~\bibnamefont {Bousquet}}, \bibinfo {author} {\bibfnamefont {Y.~M.}\ \bibnamefont {Blanter}}, \bibinfo {author} {\bibfnamefont {R.~V.}\ \bibnamefont {Mikhaylovskiy}}, \bibinfo {author} {\bibfnamefont {A.~V.}\ \bibnamefont {Kimel}},\ and\ \bibinfo {author} {\bibfnamefont {A.}~\bibnamefont {Caviglia}},\ }\bibfield  {title} {\bibinfo {title} {Ultrafast control of magnetic interactions via light-driven phonons},\ }\href@noop {} {\bibfield  {journal} {\bibinfo  {journal} {Nature materials}\ }\textbf {\bibinfo {volume} {20}},\ \bibinfo {pages} {607} (\bibinfo {year} {2021}{\natexlab{a}})}\BibitemShut {NoStop}%
\bibitem [{\citenamefont {Afanasiev}\ \emph {et~al.}(2021{\natexlab{b}})\citenamefont {Afanasiev}, \citenamefont {Hortensius}, \citenamefont {Matthiesen}, \citenamefont {Ma{\~n}as-Valero}, \citenamefont {{\v{S}}i{\v{s}}kins}, \citenamefont {Lee}, \citenamefont {Lesne}, \citenamefont {van Der~Zant}, \citenamefont {Steeneken}, \citenamefont {Ivanov} \emph {et~al.}}]{afanasiev2021controlling}%
  \BibitemOpen
  \bibfield  {author} {\bibinfo {author} {\bibfnamefont {D.}~\bibnamefont {Afanasiev}}, \bibinfo {author} {\bibfnamefont {J.~R.}\ \bibnamefont {Hortensius}}, \bibinfo {author} {\bibfnamefont {M.}~\bibnamefont {Matthiesen}}, \bibinfo {author} {\bibfnamefont {S.}~\bibnamefont {Ma{\~n}as-Valero}}, \bibinfo {author} {\bibfnamefont {M.}~\bibnamefont {{\v{S}}i{\v{s}}kins}}, \bibinfo {author} {\bibfnamefont {M.}~\bibnamefont {Lee}}, \bibinfo {author} {\bibfnamefont {E.}~\bibnamefont {Lesne}}, \bibinfo {author} {\bibfnamefont {H.~S.}\ \bibnamefont {van Der~Zant}}, \bibinfo {author} {\bibfnamefont {P.~G.}\ \bibnamefont {Steeneken}}, \bibinfo {author} {\bibfnamefont {B.~A.}\ \bibnamefont {Ivanov}}, \emph {et~al.},\ }\bibfield  {title} {\bibinfo {title} {Controlling the anisotropy of a van der waals antiferromagnet with light},\ }\href@noop {} {\bibfield  {journal} {\bibinfo  {journal} {Science advances}\ }\textbf {\bibinfo {volume} {7}},\ \bibinfo {pages} {eabf3096} (\bibinfo {year} {2021}{\natexlab{b}})}\BibitemShut
  {NoStop}%
\bibitem [{\citenamefont {Stupakiewicz}\ \emph {et~al.}(2021)\citenamefont {Stupakiewicz}, \citenamefont {Davies}, \citenamefont {Szerenos}, \citenamefont {Afanasiev}, \citenamefont {Rabinovich}, \citenamefont {Boris}, \citenamefont {Caviglia}, \citenamefont {Kimel},\ and\ \citenamefont {Kirilyuk}}]{stupakiewicz2021ultrafast}%
  \BibitemOpen
  \bibfield  {author} {\bibinfo {author} {\bibfnamefont {A.}~\bibnamefont {Stupakiewicz}}, \bibinfo {author} {\bibfnamefont {C.~S.}\ \bibnamefont {Davies}}, \bibinfo {author} {\bibfnamefont {K.}~\bibnamefont {Szerenos}}, \bibinfo {author} {\bibfnamefont {D.}~\bibnamefont {Afanasiev}}, \bibinfo {author} {\bibfnamefont {K.~S.}\ \bibnamefont {Rabinovich}}, \bibinfo {author} {\bibfnamefont {A.~V.}\ \bibnamefont {Boris}}, \bibinfo {author} {\bibfnamefont {A.}~\bibnamefont {Caviglia}}, \bibinfo {author} {\bibfnamefont {A.~V.}\ \bibnamefont {Kimel}},\ and\ \bibinfo {author} {\bibfnamefont {A.}~\bibnamefont {Kirilyuk}},\ }\bibfield  {title} {\bibinfo {title} {Ultrafast phononic switching of magnetization},\ }\href@noop {} {\bibfield  {journal} {\bibinfo  {journal} {Nature Physics}\ }\textbf {\bibinfo {volume} {17}},\ \bibinfo {pages} {489} (\bibinfo {year} {2021})}\BibitemShut {NoStop}%
\bibitem [{\citenamefont {Hansen}\ \emph {et~al.}(1977)\citenamefont {Hansen}, \citenamefont {Tolksdorf},\ and\ \citenamefont {Krishnan}}]{hansen1977anisotropy}%
  \BibitemOpen
  \bibfield  {author} {\bibinfo {author} {\bibfnamefont {P.}~\bibnamefont {Hansen}}, \bibinfo {author} {\bibfnamefont {W.}~\bibnamefont {Tolksdorf}},\ and\ \bibinfo {author} {\bibfnamefont {R.}~\bibnamefont {Krishnan}},\ }\bibfield  {title} {\bibinfo {title} {Anisotropy and magnetostriction of cobalt-substituted yttrium iron garnet},\ }\href@noop {} {\bibfield  {journal} {\bibinfo  {journal} {Physical Review B}\ }\textbf {\bibinfo {volume} {16}},\ \bibinfo {pages} {3973} (\bibinfo {year} {1977})}\BibitemShut {NoStop}%
\bibitem [{\citenamefont {Frej}\ \emph {et~al.}(2023{\natexlab{a}})\citenamefont {Frej}, \citenamefont {Davies}, \citenamefont {Kirilyuk},\ and\ \citenamefont {Stupakiewicz}}]{frej2023laser}%
  \BibitemOpen
  \bibfield  {author} {\bibinfo {author} {\bibfnamefont {A.}~\bibnamefont {Frej}}, \bibinfo {author} {\bibfnamefont {C.}~\bibnamefont {Davies}}, \bibinfo {author} {\bibfnamefont {A.}~\bibnamefont {Kirilyuk}},\ and\ \bibinfo {author} {\bibfnamefont {A.}~\bibnamefont {Stupakiewicz}},\ }\bibfield  {title} {\bibinfo {title} {Laser-induced excitation and decay of coherent optical phonon modes in an iron garnet},\ }\href@noop {} {\bibfield  {journal} {\bibinfo  {journal} {Journal of Magnetism and Magnetic Materials}\ }\textbf {\bibinfo {volume} {568}},\ \bibinfo {pages} {170416} (\bibinfo {year} {2023}{\natexlab{a}})}\BibitemShut {NoStop}%
\bibitem [{\citenamefont {Frej}\ \emph {et~al.}(2023{\natexlab{b}})\citenamefont {Frej}, \citenamefont {Davies}, \citenamefont {Kirilyuk},\ and\ \citenamefont {Stupakiewicz}}]{frej2023phonon}%
  \BibitemOpen
  \bibfield  {author} {\bibinfo {author} {\bibfnamefont {A.}~\bibnamefont {Frej}}, \bibinfo {author} {\bibfnamefont {C.}~\bibnamefont {Davies}}, \bibinfo {author} {\bibfnamefont {A.}~\bibnamefont {Kirilyuk}},\ and\ \bibinfo {author} {\bibfnamefont {A.}~\bibnamefont {Stupakiewicz}},\ }\bibfield  {title} {\bibinfo {title} {Phonon-induced magnetization dynamics in co-doped iron garnets},\ }\href@noop {} {\bibfield  {journal} {\bibinfo  {journal} {Applied Physics Letters}\ }\textbf {\bibinfo {volume} {123}} (\bibinfo {year} {2023}{\natexlab{b}})}\BibitemShut {NoStop}%
\bibitem [{\citenamefont {Marysko}(1994)}]{marysko1994anisotropy}%
  \BibitemOpen
  \bibfield  {author} {\bibinfo {author} {\bibfnamefont {M.}~\bibnamefont {Marysko}},\ }\bibfield  {title} {\bibinfo {title} {Anisotropy and {FMR} in cobalt doped {YIG} films},\ }\href@noop {} {\bibfield  {journal} {\bibinfo  {journal} {IEEE transactions on magnetics}\ }\textbf {\bibinfo {volume} {30}},\ \bibinfo {pages} {978} (\bibinfo {year} {1994})}\BibitemShut {NoStop}%
\bibitem [{\citenamefont {Mary{\v{s}}ko}(1995)}]{maryvsko1995cubic}%
  \BibitemOpen
  \bibfield  {author} {\bibinfo {author} {\bibfnamefont {M.}~\bibnamefont {Mary{\v{s}}ko}},\ }\bibfield  {title} {\bibinfo {title} {Cubic anisotropy in cobalt-doped {YIG} films},\ }\href@noop {} {\bibfield  {journal} {\bibinfo  {journal} {Journal of magnetism and magnetic materials}\ }\textbf {\bibinfo {volume} {140}},\ \bibinfo {pages} {2115} (\bibinfo {year} {1995})}\BibitemShut {NoStop}%
\bibitem [{\citenamefont {Tekielak}\ \emph {et~al.}(1997)\citenamefont {Tekielak}, \citenamefont {Andr{\"a}}, \citenamefont {Maziewski},\ and\ \citenamefont {Taubert}}]{tekielak1997magnetic}%
  \BibitemOpen
  \bibfield  {author} {\bibinfo {author} {\bibfnamefont {M.}~\bibnamefont {Tekielak}}, \bibinfo {author} {\bibfnamefont {W.}~\bibnamefont {Andr{\"a}}}, \bibinfo {author} {\bibfnamefont {A.}~\bibnamefont {Maziewski}},\ and\ \bibinfo {author} {\bibfnamefont {J.}~\bibnamefont {Taubert}},\ }\bibfield  {title} {\bibinfo {title} {Magnetic anisotropy and phase transitions in {C}o-doped yttrium iron garnet films},\ }\href@noop {} {\bibfield  {journal} {\bibinfo  {journal} {Le Journal de Physique IV}\ }\textbf {\bibinfo {volume} {7}},\ \bibinfo {pages} {C1} (\bibinfo {year} {1997})}\BibitemShut {NoStop}%
\bibitem [{\citenamefont {Hubert}\ and\ \citenamefont {Sch{\"a}fer}(2008)}]{hubert2008magnetic}%
  \BibitemOpen
  \bibfield  {author} {\bibinfo {author} {\bibfnamefont {A.}~\bibnamefont {Hubert}}\ and\ \bibinfo {author} {\bibfnamefont {R.}~\bibnamefont {Sch{\"a}fer}},\ }\href@noop {} {\emph {\bibinfo {title} {Magnetic domains: the analysis of magnetic microstructures}}}\ (\bibinfo  {publisher} {Springer Science \& Business Media},\ \bibinfo {year} {2008})\BibitemShut {NoStop}%
\bibitem [{\citenamefont {Oepts}\ \emph {et~al.}(1995)\citenamefont {Oepts}, \citenamefont {Van~der Meer},\ and\ \citenamefont {Van~Amersfoort}}]{oepts1995free}%
  \BibitemOpen
  \bibfield  {author} {\bibinfo {author} {\bibfnamefont {D.}~\bibnamefont {Oepts}}, \bibinfo {author} {\bibfnamefont {A.~F.~G.}\ \bibnamefont {Van~der Meer}},\ and\ \bibinfo {author} {\bibfnamefont {P.~W.}\ \bibnamefont {Van~Amersfoort}},\ }\bibfield  {title} {\bibinfo {title} {The free-electron-laser user facility {FELIX}},\ }\href@noop {} {\bibfield  {journal} {\bibinfo  {journal} {Infrared physics \& technology}\ }\textbf {\bibinfo {volume} {36}},\ \bibinfo {pages} {297} (\bibinfo {year} {1995})}\BibitemShut {NoStop}%
\bibitem [{\citenamefont {Janssen}\ \emph {et~al.}(2022)\citenamefont {Janssen}, \citenamefont {Davies}, \citenamefont {Gidding}, \citenamefont {Chernyy}, \citenamefont {Bakker},\ and\ \citenamefont {Kirilyuk}}]{janssen2022cavity}%
  \BibitemOpen
  \bibfield  {author} {\bibinfo {author} {\bibfnamefont {T.}~\bibnamefont {Janssen}}, \bibinfo {author} {\bibfnamefont {C.}~\bibnamefont {Davies}}, \bibinfo {author} {\bibfnamefont {M.}~\bibnamefont {Gidding}}, \bibinfo {author} {\bibfnamefont {V.}~\bibnamefont {Chernyy}}, \bibinfo {author} {\bibfnamefont {J.}~\bibnamefont {Bakker}},\ and\ \bibinfo {author} {\bibfnamefont {A.}~\bibnamefont {Kirilyuk}},\ }\bibfield  {title} {\bibinfo {title} {Cavity-dumping a single infrared pulse from a free-electron laser for two-color pump--probe experiments},\ }\href@noop {} {\bibfield  {journal} {\bibinfo  {journal} {Review of Scientific Instruments}\ }\textbf {\bibinfo {volume} {93}} (\bibinfo {year} {2022})}\BibitemShut {NoStop}%
\bibitem [{\citenamefont {Zalewski}\ \emph {et~al.}(2025)\citenamefont {Zalewski}, \citenamefont {Kirilyuk},\ and\ \citenamefont {Davies}}]{zalewski2025direct}%
  \BibitemOpen
  \bibfield  {author} {\bibinfo {author} {\bibfnamefont {T.}~\bibnamefont {Zalewski}}, \bibinfo {author} {\bibfnamefont {A.}~\bibnamefont {Kirilyuk}},\ and\ \bibinfo {author} {\bibfnamefont {C.}~\bibnamefont {Davies}},\ }\bibfield  {title} {\bibinfo {title} {Direct visualization of ultrafast and inhomogeneous self-focusing effects in semiconductors},\ }\href@noop {} {\bibfield  {journal} {\bibinfo  {journal} {Physical Review B}\ }\textbf {\bibinfo {volume} {112}},\ \bibinfo {pages} {L220302} (\bibinfo {year} {2025})}\BibitemShut {NoStop}%
\bibitem [{\citenamefont {Clark}\ and\ \citenamefont {Strakna}(1961)}]{clark1961elastic}%
  \BibitemOpen
  \bibfield  {author} {\bibinfo {author} {\bibfnamefont {A.}~\bibnamefont {Clark}}\ and\ \bibinfo {author} {\bibfnamefont {R.}~\bibnamefont {Strakna}},\ }\bibfield  {title} {\bibinfo {title} {Elastic constants of single-crystal yig},\ }\href@noop {} {\bibfield  {journal} {\bibinfo  {journal} {Journal of Applied Physics}\ }\textbf {\bibinfo {volume} {32}},\ \bibinfo {pages} {1172} (\bibinfo {year} {1961})}\BibitemShut {NoStop}%
\bibitem [{\citenamefont {Zhou}\ \emph {et~al.}(2011)\citenamefont {Zhou}, \citenamefont {Li}, \citenamefont {Nellis}, \citenamefont {Wang}, \citenamefont {Li}, \citenamefont {He},\ and\ \citenamefont {Wu}}]{zhou2011pressure}%
  \BibitemOpen
  \bibfield  {author} {\bibinfo {author} {\bibfnamefont {X.}~\bibnamefont {Zhou}}, \bibinfo {author} {\bibfnamefont {J.}~\bibnamefont {Li}}, \bibinfo {author} {\bibfnamefont {W.~J.}\ \bibnamefont {Nellis}}, \bibinfo {author} {\bibfnamefont {X.}~\bibnamefont {Wang}}, \bibinfo {author} {\bibfnamefont {J.}~\bibnamefont {Li}}, \bibinfo {author} {\bibfnamefont {H.}~\bibnamefont {He}},\ and\ \bibinfo {author} {\bibfnamefont {Q.}~\bibnamefont {Wu}},\ }\bibfield  {title} {\bibinfo {title} {Pressure-dependent hugoniot elastic limit of gd3ga5o12 single crystals},\ }\href@noop {} {\bibfield  {journal} {\bibinfo  {journal} {Journal of Applied Physics}\ }\textbf {\bibinfo {volume} {109}} (\bibinfo {year} {2011})}\BibitemShut {NoStop}%
\bibitem [{\citenamefont {Vansteenkiste}\ \emph {et~al.}(2014)\citenamefont {Vansteenkiste}, \citenamefont {Leliaert}, \citenamefont {Dvornik}, \citenamefont {Helsen}, \citenamefont {Garcia-Sanchez},\ and\ \citenamefont {Van~Waeyenberge}}]{vansteenkiste2014design}%
  \BibitemOpen
  \bibfield  {author} {\bibinfo {author} {\bibfnamefont {A.}~\bibnamefont {Vansteenkiste}}, \bibinfo {author} {\bibfnamefont {J.}~\bibnamefont {Leliaert}}, \bibinfo {author} {\bibfnamefont {M.}~\bibnamefont {Dvornik}}, \bibinfo {author} {\bibfnamefont {M.}~\bibnamefont {Helsen}}, \bibinfo {author} {\bibfnamefont {F.}~\bibnamefont {Garcia-Sanchez}},\ and\ \bibinfo {author} {\bibfnamefont {B.}~\bibnamefont {Van~Waeyenberge}},\ }\bibfield  {title} {\bibinfo {title} {The design and verification of {MuMax3}},\ }\href@noop {} {\bibfield  {journal} {\bibinfo  {journal} {AIP advances}\ }\textbf {\bibinfo {volume} {4}} (\bibinfo {year} {2014})}\BibitemShut {NoStop}%
\bibitem [{\citenamefont {Klingler}\ \emph {et~al.}(2014)\citenamefont {Klingler}, \citenamefont {Chumak}, \citenamefont {Mewes}, \citenamefont {Khodadadi}, \citenamefont {Mewes}, \citenamefont {Dubs}, \citenamefont {Surzhenko}, \citenamefont {Hillebrands},\ and\ \citenamefont {Conca}}]{klingler2014measurements}%
  \BibitemOpen
  \bibfield  {author} {\bibinfo {author} {\bibfnamefont {S.}~\bibnamefont {Klingler}}, \bibinfo {author} {\bibfnamefont {A.~V.}\ \bibnamefont {Chumak}}, \bibinfo {author} {\bibfnamefont {T.}~\bibnamefont {Mewes}}, \bibinfo {author} {\bibfnamefont {B.}~\bibnamefont {Khodadadi}}, \bibinfo {author} {\bibfnamefont {C.}~\bibnamefont {Mewes}}, \bibinfo {author} {\bibfnamefont {C.}~\bibnamefont {Dubs}}, \bibinfo {author} {\bibfnamefont {O.}~\bibnamefont {Surzhenko}}, \bibinfo {author} {\bibfnamefont {B.}~\bibnamefont {Hillebrands}},\ and\ \bibinfo {author} {\bibfnamefont {A.}~\bibnamefont {Conca}},\ }\bibfield  {title} {\bibinfo {title} {Measurements of the exchange stiffness of {YIG} films using broadband ferromagnetic resonance techniques},\ }\href@noop {} {\bibfield  {journal} {\bibinfo  {journal} {Journal of Physics D: Applied Physics}\ }\textbf {\bibinfo {volume} {48}},\ \bibinfo {pages} {015001} (\bibinfo {year} {2014})}\BibitemShut {NoStop}%
\bibitem [{\citenamefont {Stupakiewicz}\ \emph {et~al.}(2017)\citenamefont {Stupakiewicz}, \citenamefont {Szerenos}, \citenamefont {Afanasiev}, \citenamefont {Kirilyuk},\ and\ \citenamefont {Kimel}}]{stupakiewicz2017ultrafast}%
  \BibitemOpen
  \bibfield  {author} {\bibinfo {author} {\bibfnamefont {A.}~\bibnamefont {Stupakiewicz}}, \bibinfo {author} {\bibfnamefont {K.}~\bibnamefont {Szerenos}}, \bibinfo {author} {\bibfnamefont {D.}~\bibnamefont {Afanasiev}}, \bibinfo {author} {\bibfnamefont {A.}~\bibnamefont {Kirilyuk}},\ and\ \bibinfo {author} {\bibfnamefont {A.~V.}\ \bibnamefont {Kimel}},\ }\bibfield  {title} {\bibinfo {title} {Ultrafast nonthermal photo-magnetic recording in a transparent medium},\ }\href@noop {} {\bibfield  {journal} {\bibinfo  {journal} {Nature}\ }\textbf {\bibinfo {volume} {542}},\ \bibinfo {pages} {71} (\bibinfo {year} {2017})}\BibitemShut {NoStop}%
\bibitem [{\citenamefont {Smith}\ and\ \citenamefont {Jones}(1963)}]{smith1963magnetostriction}%
  \BibitemOpen
  \bibfield  {author} {\bibinfo {author} {\bibfnamefont {A.}~\bibnamefont {Smith}}\ and\ \bibinfo {author} {\bibfnamefont {R.}~\bibnamefont {Jones}},\ }\bibfield  {title} {\bibinfo {title} {Magnetostriction constants from ferromagnetic resonance},\ }\href@noop {} {\bibfield  {journal} {\bibinfo  {journal} {Journal of Applied Physics}\ }\textbf {\bibinfo {volume} {34}},\ \bibinfo {pages} {1283} (\bibinfo {year} {1963})}\BibitemShut {NoStop}%
\bibitem [{\citenamefont {Gidding}\ \emph {et~al.}(2023)\citenamefont {Gidding}, \citenamefont {Janssen}, \citenamefont {Davies},\ and\ \citenamefont {Kirilyuk}}]{gidding2023dynamic}%
  \BibitemOpen
  \bibfield  {author} {\bibinfo {author} {\bibfnamefont {M.}~\bibnamefont {Gidding}}, \bibinfo {author} {\bibfnamefont {T.}~\bibnamefont {Janssen}}, \bibinfo {author} {\bibfnamefont {C.~S.}\ \bibnamefont {Davies}},\ and\ \bibinfo {author} {\bibfnamefont {A.}~\bibnamefont {Kirilyuk}},\ }\bibfield  {title} {\bibinfo {title} {Dynamic self-organisation and pattern formation by magnon-polarons},\ }\href@noop {} {\bibfield  {journal} {\bibinfo  {journal} {Nature Communications}\ }\textbf {\bibinfo {volume} {14}},\ \bibinfo {pages} {2208} (\bibinfo {year} {2023})}\BibitemShut {NoStop}%
\bibitem [{\citenamefont {Lourens}\ \emph {et~al.}(2026)\citenamefont {Lourens}, \citenamefont {Kwaaitaal}, \citenamefont {Davies},\ and\ \citenamefont {Kirilyuk}}]{lourens2026ferroelastic}%
  \BibitemOpen
  \bibfield  {author} {\bibinfo {author} {\bibfnamefont {D.~G.}\ \bibnamefont {Lourens}}, \bibinfo {author} {\bibfnamefont {M.}~\bibnamefont {Kwaaitaal}}, \bibinfo {author} {\bibfnamefont {C.~S.}\ \bibnamefont {Davies}},\ and\ \bibinfo {author} {\bibfnamefont {A.}~\bibnamefont {Kirilyuk}},\ }\bibfield  {title} {\bibinfo {title} {Ferroelastic domain switching by ultrafast photoinduced strain},\ }\href@noop {} {\bibfield  {journal} {\bibinfo  {journal} {arXiv preprint arXiv:2609.02321}\ } (\bibinfo {year} {2026})}\BibitemShut {NoStop}%
\end{thebibliography}
\end{document}